# Digital Audio Tampering Detection Based on ENF Spatio-temporal Features Representation Learning

Chunyan Zeng,[1] Shuai Kong,[1] Zhifeng Wang,[2] Xiangkui Wan,[1] and Yunfan Chen[1]

[1] Hubei Key Laboratory for High-efficiency Utilization of Solar Energy and Operation Control of Energy Storage System, Hubei University of Technology, Nanli Road 28, 430068, Wuhan, China.
[2] Department of Digital Media Technology, Central China Normal University, Luoyu Road 152, 430079, Wuhan, China.

Correspondence should be addressed to Zhifeng Wang; zfwang@ccnu.edu.cn

## Abstract

Most digital audio tampering detection methods based on electrical network frequency (ENF) only utilize the static spatial information of ENF, ignoring the variation of ENF in time series, which limit the ability of ENF feature representation and reduce the accuracy of tampering detection. This paper proposes a new method for digital audio tampering detection based on ENF spatio-temporal features representation learning. A parallel spatio-temporal network model is constructed using CNN and BiLSTM, which deeply extracts ENF spatial feature information and ENF temporal feature information to enhance the feature representation capability to improve the tampering detection accuracy. In order to extract the spatial and temporal features of the ENF, this paper firstly uses digital audio high-precision Discrete Fourier Transform analysis to extract the phase sequences of the ENF. The unequal phase series is divided into frames by adaptive frame shifting to obtain feature matrices of the same size to represent the spatial features of the ENF. At the same time, the phase sequences are divided into frames based on ENF time changes information to represent the temporal features of the ENF. Then deep spatial and temporal features are further extracted using CNN and BiLSTM respectively, and an attention mechanism is used to adaptively assign weights to the deep spatial and temporal features to obtain spatio-temporal features with stronger representation capability. Finally, the deep neural network is used to determine whether the audio has been tampered with. The experimental results show that the proposed method improves the accuracy by 2.12%-7.12% compared with state-of-the-art methods under the public database Carioca, New Spanish.

## 1. Introduction

With the rapid development of information technology, digital audio has become an essential part of the communication of public information. Digital audio, such as telephone recordings, voice messages, and music files, can be found everywhere in our daily life [1]. Due to the low threshold and powerful operation of existing audio editing software, digital audio tampering can be easily accomplished by a normal user without any expertise in audio processing [2]. Therefore, unscrupulous individuals may use tampering with digital audio to try to escape the law and even cause damage to society. In addition, digital audio tampering fragments are often milliseconds in duration and such subtle forgeries are difficult to identify [1]. As a result, there is a growing demand for digital audio forensic methods, particularly in the areas of judicial forensics, scientific discovery, and commercial applications [3, 4].

There are two main types of forensic techniques for digital audio: active forensics and passive forensics [1]. Digital audio active forensics focuses on determining the authenticity or integrity of audio by detecting whether a pre-embedded digital signature or digital watermark has been compromised. However, in practice, most audio signals are recorded without a pre-embedded watermark or signature, so digital audio active forensics has limitations in its application. Digital audio tampering passive detection means relying only on the characteristics of the digital audio itself to discern the authenticity and integrity of the digital audio without adding any information. This passive detection of digital audio tampering is more practical for audio forensics in complex environments, so this paper focuses on the passive detection of digital audio tampering.

In recent years, there have been many research results in the field of passive detection of digital audio tampering. These researches mainly focus on the selection of audio features such as discrepancy information of background noise [5,6,7,8,9], spectrograms of audio content [10,11], pitch [12,13,14], resonance peaks [14], electrical network frequency (ENF) discrepancy information [15], ENF harmonic signals [16] and ENF phase and frequency information [17,18,19]. The ENF has the characteristic of fluctuating randomly at a nominal frequency (50 or 60Hz), which is automatically embedded in the audio when recorded and has certain stability and uniqueness [20]. Therefore, the ENF is widely used for digital audio tampering detection. Most of the methods based on the ENF digital audio tampering detection extract the ENF feature information and achieve tampering detection by means of classification algorithms. However, in terms of feature selection, the ability to represent information related to digital audio tampering detection in traditional ENF features is not strong; in terms of classification algorithm selection, most existing methods use traditional machine learning algorithms. However, traditional machine learning algorithms tend to ignore the interconnectedness of features in practice and do not have a significant advantage in the reinforcement of important features.

To address the problems of weak feature representation and poor classification accuracy, this paper proposes a method for digital audio tampering detection based on ENF spatio-temporal features representation learning. For the extraction of spatial and temporal features of the ENF, this paper firstly extracts the phase sequences of the ENF using a high-precision Discrete Fourier Transform (DFT) analysis method. The unequal phase sequences are split into frames by adaptive frame shifting to obtain a matrix of the same size to represent the spatial features of the ENF. At the same time, the phase sequences are split into frames according to the ENF timing change information to represent the temporal features of the ENF. For the construction of a parallel CNN-BiLSTM network model, the model consists of four main components: a deep spatial feature extraction module, a deep temporal feature extraction module, a spatio-temporal feature fusion module, and a classification module. First, in the deep spatial feature extraction module, we exploit the excellent spatial representation capability of the convolution neural network (CNN) to extract deep spatial features. While in the deep temporal feature extraction module, we exploit the ability of Bidirectional Long Short-Term Memory (BiLSTM) to handle temporal signals well to extract deep temporal features. Then, in the spatio-temporal feature fusion module, we use an attention mechanism to adaptively assign weights to deep spatial and temporal features to obtain the fused spatio-temporal features. Finally, we adjudicate whether the audio has been tampered with through the MLP network.

The main contributions made in this paper are as follows:

1. Based on the extraction of high precision ENF phase sequences, this paper divides frames by adaptive frame shifting to obtain phase feature matrices of the same size to represent the spatial features of the ENF. At the same time, the frames are divided according to the ENF timing change information to represent the temporal features of the ENF. The tampering information in the ENF is mined in multiple dimensions to enhance the representation capability of the features.
2. We use the spatial representation capability of the CNN and the time series processing capability of the BiLSTM networks to extract deep spatial features and deep temporal features in ENF, and use the attention mechanism to adaptively assign the weights of spatial and temporal information to achieve the fusion of spatio-temporal features. The fused spatio-temporal features complement each other and improve the detection accuracy.

The remainder of the paper is organized as follows: Section 2 presents related work that has already been done. Section 3 describes the framework proposed in this paper. Section 4 presents the dataset used to evaluate the performance of the framework, details of the specific experimental setup, and comparison experiments. Finally, the paper concludes in Section 5 and lists directions for future work.

## 2. Related work

The problem to be solved in digital audio tampering passive detection research is to determine the authenticity and integrity of the audio [20]. There are three types of analysis from the perspective of the constituent components of digital audio: the first is the analysis based on the background noise of digital audio; the second is the analysis based on the content features of digital audio; the third is the analysis based on the ENF in digital audio.

### 2.1 Analysis based on background noise

During digital audio recording, the audio is certain to contain information about the background noise of the current environment. The background noise information of recorded audio has short-term stability, and tampering operations cause discontinuities in the background noise in the audio, so the background noise can be used as a basis for tampering detection.

Tampering detection based on background noise can be divided into two steps, firstly the separation of the noise from the audio content and secondly the discrepancy analysis of the noise features. X. Meng et al. [5] determined whether tampering operations were present in the audio by comparing the similarity of the background noise variance of each syllable. However, information on the variance of syllable variance under weak noise conditions was insignificant and had a significant impact on tampering detection accuracy. To improve the accuracy of tampering detection under weak noise conditions, X. Lin et al. [6] used spectral phase reconstruction to counteract the effect of noise by analyzing the higher order statistics of the spectral phase residuals and the baseband phase correlation between two adjacent turbid bands to determine whether the audio was tampered with. In addition, D. Yan et al. [7] used a change-point detection algorithm based on a parameter-optimized noise estimation algorithm to achieve tamper region localization for noisy signals. C. Zeng et al [27] proposed a parallel neural network structure. Firstly, GSV features and MFCC features are input to the spatial information extraction network and temporal information extraction network respectively, then the weights of spatial and temporal information are assigned by the attention mechanism, and finally, the classification of device noise is achieved by Deep Neural Networks (DNN).

Although there have been many research results based on the analysis of background noise [5, 6, 7], there are still some limitations. One is that the complexity of the actual environment is often difficult to predict, and the other is that the noise in the sampling of very short audio separation from audio is difficult. It is worth noting that the selection of the optimal set of special features to represent the background noise in a consistent analysis of the noise is still a question worth exploring.

**2.2 Analysis based on audio content features**

When digital audio content is tampered with, the correlation between frames is weakened and the features of the audio content change. Therefore, we can determine whether digital audio has been tampered with based on the discrepancy in audio content features. S. Saleem et al. [11] processed the audio by Short Time Fourier Transform (STFT) and Modified Discrete Cosine Transform (MDCT) to obtain a sound spectrum map on a logarithmic scale and fed its plotted pattern to the CNN to identify the base-transformed audio. In addition, C. Li et al. [12] used the fundamental tone sequences as the audio features and used a differential comparison algorithm to achieve tamper detection. However, C. Li et al. focused only on the fundamental tones in the audio content, but ignored the information on the variability of the resonance peaks in the audio content. On this basis, Q. Yan et al [13] extracted the base tone sequences and the first two resonant peak sequences of the audio segment as feature sets. The Dynamic Time Warping (DTW) algorithm was used to calculate the similarity of each feature set and compare it with a threshold value to detect and locate tampering operations in the audio. However, there is an empirical behaviour of threshold selection in [12, 13] that fails to allow automated tamper detection.

**2.3 Analysis based on ENF**

When the recording device is powered by the electric network, the ENF will be automatically embedded in the recording file. Since the ENF signal has certain stability and uniqueness and shows reliable discrimination in audio tampering detection studies [21], the technique based on ENF analysis has been widely used in the field of digital audio tampering detection [20].

Based on these characteristics, researchers usually extract the ENF signal from digital audio and analyze the ENF with consistency or regularity. D. Rodríguez et al. [17] estimated the ENF phase variation by high-precision DFT analysis. In addition, P. Reis et al. [18] proposed to measure ENF fluctuations based on the Estimation of Signal Parameters via Rotational Invariance Techniques (ESPRIT) estimation of phase peak features for the case of tampered signal ENF phase discontinuities and to automatically detect sudden changes in ENF using Support Vector Machine (SVM). However, utilizing only the tampering information in the phase features of the ENF had limitations in terms of feature representation capability. On this basis, Z. Wang et al. [19] extracted the phase features of the ENF component (ENFC) based on $DFT^0$ and $DFT^1$, extracted the instantaneous frequency features of ENFC based on Hilbert transform, and used the SVM classifier to determine whether had been tampered with. Besides utilizing ENF phase and frequency features, M. Mao et al. [22] used multi-signal classification, Hilbert linear prediction, and Welch's algorithm to extract ENF features from audio signals, and passed the extracted features through CNN for detection. M. Sarkar et al. [23] decomposed the extracted ENF into low outlier and high outlier frequency segments, and then used statistical and signal processing analysis to determine the potential feature vectors of the ENF segments, and finally, the SVM classifier was used for validation.

For the selection of features, [17, 18, 19, 22] extracted ENF static spatial information to achieve audio tamper detection. But these methods result in the loss of ENF timing information so that the feature representation was weak. In terms of classification algorithms, [17, 18, 19, 23] used the audiovisual analysis or traditional machine learning methods, which fail to reinforce important features and mine deeper information, causing tampering detection to be less accurate. Since the ENF in audio fluctuates randomly with time, making full use of the temporal information of the ENF can improve the ability to represent features. On this basis, we propose a digital audio tampering detection method based on ENF spatio-temporal features representation learning. For feature selection, we extract phase sequences based on high precision $DFT^1$ and then use different frame processing to extract spatial and temporal features of the ENF for representation learning. On the parallel CNN-BiLSTM network model, we use the spatial representation capability of CNN to further extract deep spatial features, and the excellent temporal processing capability of BiLSTM to further extract deep temporal features. The attention mechanism is then used to achieve the fusion of the temporal-spatial features. Finally, tampering detection is achieved through the MLP classification network.

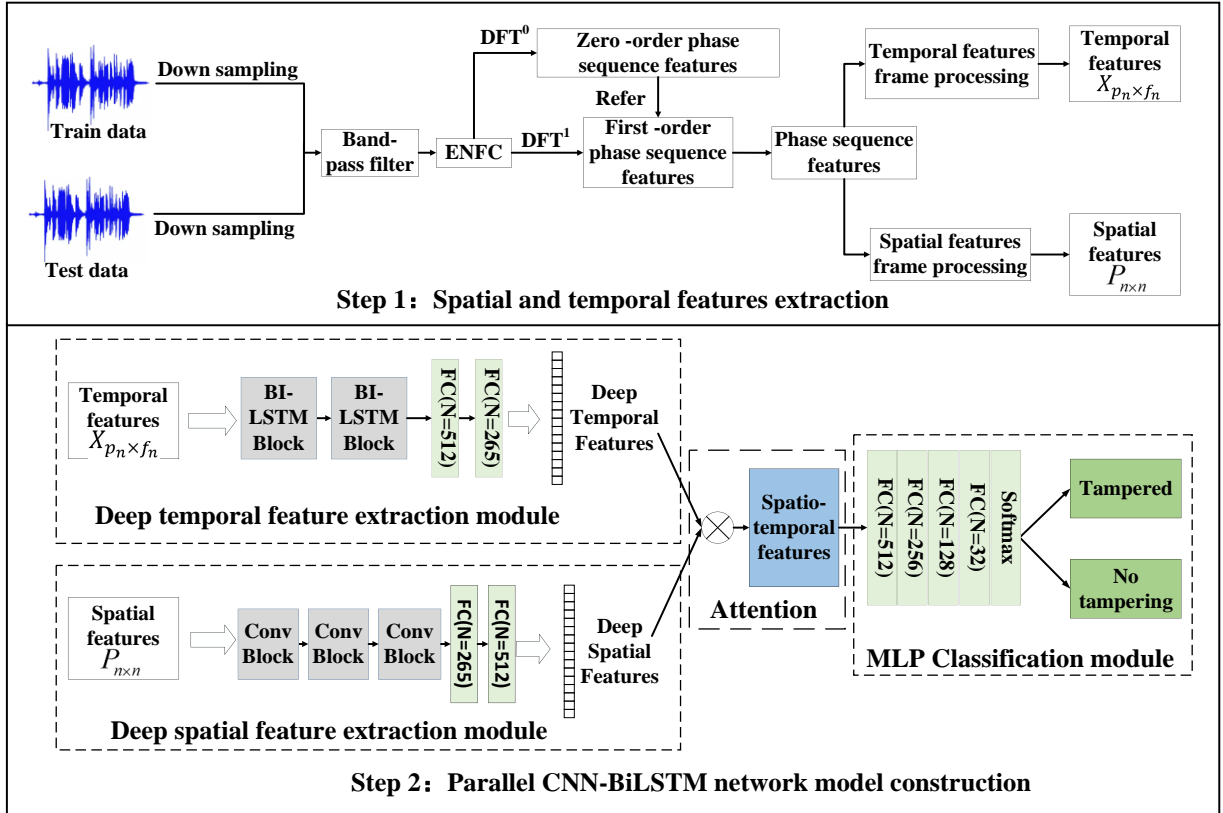

Figure 1: Framework diagram of the digital audio tampering detection system based on ENF spatio-temporal feature fusion.

## 3. Methods

We propose the digital audio tampering detection method based on ENF spatio-temporal features representation learning. The method is divided into two main parts: spatial and temporal features extraction and the construction of a parallel CNN-BiLSTM network model, as shown in Figure 1. In the extraction of spatial and temporal features, the ENFC is firstly obtained by filtering the training and test audio; secondly, the high precision phase sequence $\psi_1$ is extracted based on $DFT^1$ [19], and then different split frame algorithms are used to characterize the spatial and temporal features of the ENF respectively, as detailed in Section 3.1. In parallel CNN-BiLSTM network model construction, firstly deep spatial features and deep temporal features are extracted by parallel deep spatial feature extraction module and deep temporal feature extraction module, respectively, as detailed in sections 3.2.1 and 3.2.2;

then, the attention mechanism is used to adaptively assign weights and fuse the temporal feature information and spatial feature information of ENF to achieve the fusion of temporal-spatial features, as detailed in Section 3.2.3; finally, the MLP classification network is used to determine whether the digital audio has been tampered with, as detailed in Section 3.2.4.

### 3.1 Representation of the spatial and temporal features of the ENF

We improve the representation ability of ENF features by extracting both spatial and temporal features of ENF. Firstly, the ENF first-order phase sequence features are extracted based on the highly accurate DFT; secondly, the frames are divided by adaptive frame shifting to obtain a phase feature matrix of the same size to represent the spatial features of the ENF; at the same time, the frames are divided according to the ENF phase time variation information to represent the temporal features of the ENF.

### 3.1.1 Extracting the first-order phase features of the ENF

To extract the first-order phase features of the ENF, firstly the ENFC of the digital audio is extracted by down sampling and band-pass filtering. Then, the ENFC performed $DFT^0$ to extract the phase sequence feature $\psi_0$. Finally, $DFT^1$ is performed on the ENFC to estimate the first-order phase sequence feature $\psi_1$ of the ENF, where $DFT^k$ denotes the $DFT$ of the kth-order derivative of the signal [17]. The extraction of the first-order phase features $\psi_1$ of the ENF is divided into three parts: pre-processing of the digital audio signal, estimation of the phase sequence $\psi_0$ based on $DFT^0$, and estimation of the first-order phase sequence $\psi_1$ based on $DFT^1$.

---

**Algorithm 1**: Algorithm to extract $\psi_0, \psi_1$

**Input:** ENF component: $U_{ENFC}[n]$, Down sampling frequency: $f_d$;
**Output:** $\psi_0, \psi_1$;

1. Calculate the first derivative of $U_{ENFC}[n]$:
$$u'_{ENFC}[n] = f_d(U_{ENFC}[n] - U_{ENFC}[n-1]);$$
2. Get the windowed signal:
$$U'_N[n] = u'_{ENFC}[n]w(n),$$
$$U_N[n] = U_{ENFC}[n]w(n);$$
3. $DFT$ transform $U_N[n]$ and $U'_N[n]$ to get $U[k]$ and $U'[k]$;
4. By using the maximum value $k_{peak}$ of each frame signal as the integer index of $|U[k]|$ and $|U'[k]|$, $U[k_{peak}]$ and $U'[k_{peak}]$ can be obtained;
5. Calculate the phase information $\psi_0$:
$$\psi_0 = arg[U(k_{peak})];$$
6. Estimated frequency value: $f_{DFT^1} = \frac{1}{2\pi}\frac{DFT^1[k_{peak}]}{DFT^0[k_{peak}]}$;
7. Calculate the phase information $\psi_1$:
$$\psi_1 = arctan\left\{\frac{tan(\theta)[1-cos(\omega_0)+sin(\omega_0)]}{1-cos(\omega_0)-tan(\theta)sin(\omega_0)}\right\},$$
$$\theta \approx (k_{DFT^1} - k_{low})\frac{\theta_{high}-\theta_{low}}{k_{high}-k_{low}} + \theta_{low};$$

---

In the pre-processing stage of the digital audio signal, in order to extract the ENFC from the digital audio, firstly the digital audio signal $u[n]$ is down sampled to obtain the down sampling signal $u_d[n]$, where the down sampling frequency $f_d$ is set to 1000Hz or 1200Hz.

Then, we filter the down-sampled signal $u_d[n]$ using band-pass filtering to obtain the ENFC $U_{ENFC}[n]$ of the signal to be measured.

To obtain the 0th order phase sequence $\psi_0$, firstly we need to frame and add a window to $U_{ENFC}[n]$, where the frame length is ten standard ENF frequency periods and the frame shift is one standard ENF period. The window function uses the Hanning window $w(n)$, and $U_N[n]$ is the signal after the frame-splitting and windowing. Then, we perform a DFT for $U_N[n]$ at N points to obtain $U(k)$, and make $k_{peak}$ the integer index of the maximum of $|U(k)|$. Finally, the 0th order phase sequence $\psi_0$ of the ENF signal is $arg[U(k_{peak})]$, as shown in Algorithm 1.

In order to estimate the phase information more accurately, this paper uses $DFT^1$ to further extract the first-order phase sequence features $\psi_1$ based on $\psi_0$. Firstly, the approximate first order derivative of the ENF signal $U_{ENFC}[n]$ at point $n$ is calculated as follows:

$$u'_{ENFC}[n] = f_d(U_{ENFC}[n] - U_{ENFC}[n-1]) \tag{1}$$

Secondly, a window is added to the first order derivative signal $U'_{ENFC}[n]$. Then, we estimate the frequency values based on $U'[k_{peak}]$ as:

$$f_{DFT^1} = \frac{1}{2\pi}\frac{DFT^1[k_{peak}]}{DFT^0[k_{peak}]} \tag{2}$$

where $DFT^0[k_{peak}] = U[k_{peak}]$ and $DFT^1[k_{peak}] = F(k)U'[k_{peak}]$. $F(k)$ is a scale function. Finally, we use the $DFT^1$ method to estimate $\psi_1$ as:

$$\psi_1 = arctan\left\{\frac{tan(\theta)[1-cos(\omega_0)+sin(\omega_0)]}{1-cos(\omega_0)-tan(\theta)sin(\omega_0)}\right\} \tag{3}$$

where $\omega_0 \approx 2\pi f_{DFT^1}/f_d$. There are two possible values for the solution of $\psi_1$. In this paper, we use $\psi_0$ as a reference and choose the value of $\psi_1$ that is closest to $\psi_0$ as the final solution. We perform a linear interpolation of $U'[k]$ to obtain the value of $\theta$. The formula for calculating $\theta$ is:

$$\theta \approx (k_{DFT^1} - k_{low})\frac{\theta_{high}-\theta_{low}}{k_{high}-k_{low}} + \theta_{low} \tag{4}$$

where $k_{low} = floor[k_{DFT^1}]$ and $k_{high} = ceil[k_{DFT^1}]$. $floor[i]$ represents the largest integer less than $i$ and $ceil[j]$ represents the largest integer greater than $j$. $k_{DFT^1} = f_{DFT^1}N_{DFT}/f_d$, $f_d$ is the down sampling frequency. The process of estimating the phase sequence $\psi_1$ based on $DFT^1$ is shown in Algorithm 1.

### 3.1.2 Extraction of spatial features of ENF

When digital audio is tampered with, it can cause abrupt changes in the ENF phase characteristics [17]. In order to represent the static abrupt change information in the phase sequences, we have designed a split frame method for the static features of the phase sequences to extract the spatial feature matrices $P_{n\times n}$ of the ENF. The varying lengths of the individual samples in the dataset have caused the extracted phase sequence feature $\psi_1$ to vary in length. In order to reduce information loss during feature extraction, we have designed an adaptive frame shifting approach for frame splitting to extract the spatial features $P_{n\times n}$ of the ENF so that the audio ENF phase features $\psi_1$ become $n \times n$ matrix, where $n$ is the number of phase points per unit frame in the spatial features (the value of $n$ is determined by the longest audio data). By giving the unequal audio phase sequences an adaptive frame shift, the final spatial feature matrix obtained is of the same shape and size to facilitate automatic learning of the CNN, as shown in Algorithm 2.

---
**Algorithm 2**: Algorithm to extract spatial features $P_{n\times n}$

**Input**: Phase sequence features $\psi_1$;
**Output:** spatial features: $P_{n\times n}$;
1. Calculate the longest phase: $len(\psi_{DFT^1})$;
2. The number of phase points contained in a frame $n$:
$$n = ceil(\sqrt{X}), \text{ where } X = ceil(\sqrt{len(\psi_{DFT^1})});$$
3. **For** Phase sequence features of all audio **do**
    Calculate the frame shift: $overlap = n - ceil(\frac{X-n}{n-1})$;
    Split the frame;
    Reshape into feature matrix $P_{n\times n}$;
   **end**
4. **return** $P_{n\times n}$;

---

### 3.1.3 Extraction of temporal features of ENF

---
**Algorithm 3**: Algorithms for frame processing.

**Input:** Phase sequence features: $\psi_1$, The number of phase points contained in a frame: $p_n$;
**Output:** Temporal features of ENF: $X_{p_n \times f_n}$;
1. Calculate the length of the phase sequence: $M_{max}$;
2. Calculate the number of frames: $f_n = \frac{M_{max}}{p_n}$;
3. **For** Phase sequence features of all audio **do**
    Calculate the frame shift: $overlap = p_n - floor[\frac{length(\psi_1)}{f_n}]$;
    Split the frame;
    Reshape into feature matrix $X_{p_n \times f_n}$;
   **end**
4. **return** $X_{p_n \times f_n}$;

---

Due to the non-periodic fluctuations of the ENF, the parameters of the various information and basic characteristics of the ENF change with time. In order to represent the timing information of the ENF, this paper designs a method of splitting frames according to the way the ENF timing changes to obtain the ENF temporal features. Unlike the spatial representation in the previous section, the purpose of extracting the temporal features $X_{p_n \times f_n}$ in this section is to reduce information loss and to use recurrent neural networks to learn information about the long-term variation of the ENF phase. In the temporal features $X_{p_n \times f_n}$, $p_n$ denotes the number of phase points per unit frame and $f_n$ denotes the number of frames. Different frame length settings result in a change in the number of phase points $p_n$ in the frame length, which affects the representation of the ENF phase time sequences over a period of time. $p_n$ is an artificially set value and will be used as a variable for experiments on ENF time series representation in subsequent sections of this paper. The specific steps for ENF phase temporal features extraction are shown in Algorithm 3.

### 3.2 Construction of a parallel CNN-BiLSTM network framework

We have designed a parallel CNN-BiLSTM network model based on the spatial features $P_{n \times n}$ and temporal features $X_{p_n \times f_n}$ extracted in 3.1 to implement passive detection of digital audio tampering, as shown in Figure 2. The spatial features $P_{n \times n}$ and temporal features $X_{p_n \times f_n}$ are put into the CNN network and BiLSTM network respectively for representation learning to obtain the deep spatial and temporal features. Then, in the feature fusion module, we use the attention mechanism to fuse deep spatial and temporal features in digital audio to achieve the fusion of spatio-temporal features. The attention mechanism assigns the weights of deep spatial and temporal features in an adaptive manner so that more representational spatio-temporal features are obtained. Finally, tampering detection is achieved by the MLP neural network.

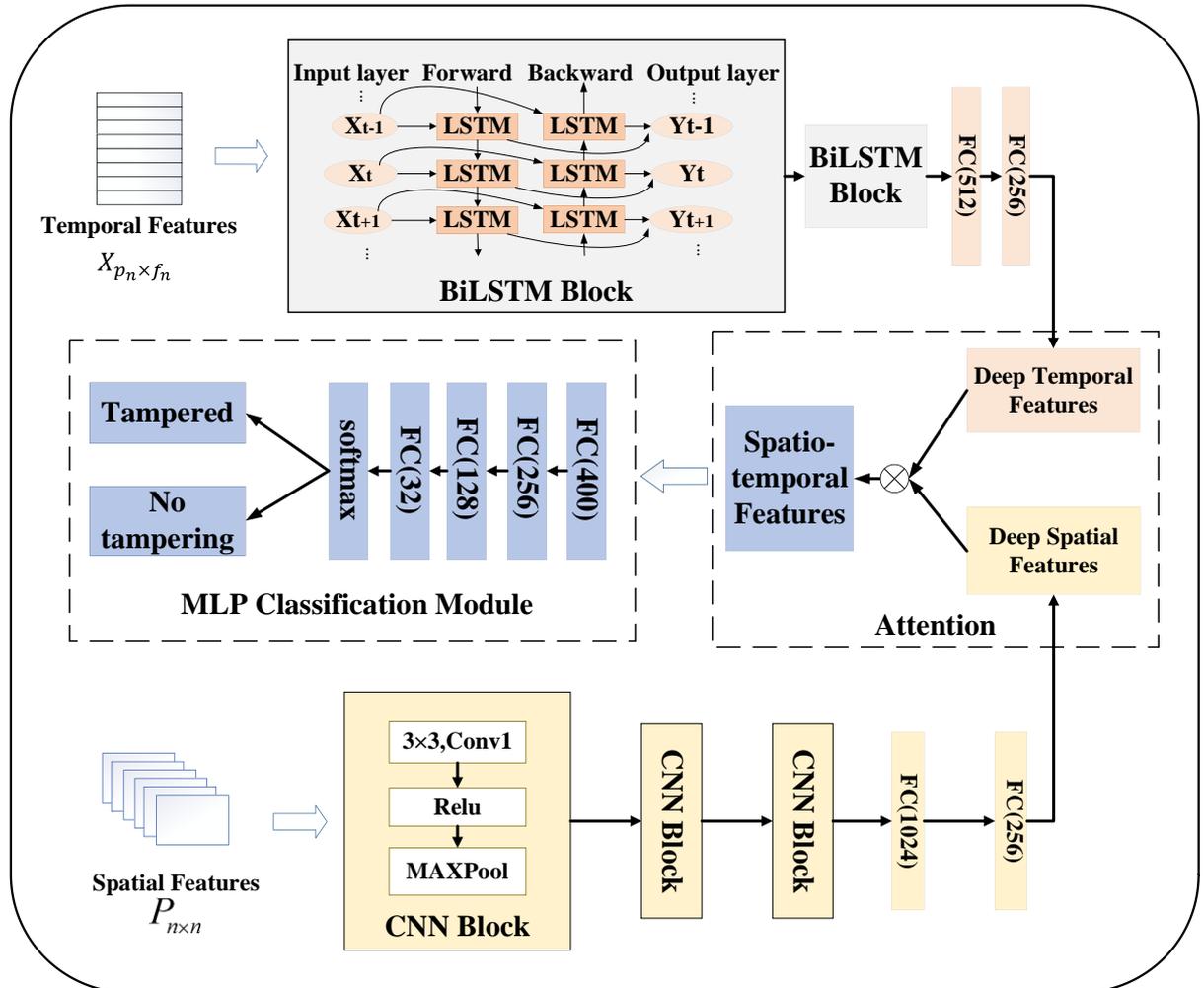

Figure 2: A parallel CNN-BiLSTM network based on spatio-temporal features fusion, the network inputs are spatial features $P_{n \times n}$ and temporal features $X_{p_n \times f_n}$.

### 3.2.1 Extraction of deep spatial features based on CNN

In this section, in order to extract the deep spatial features of the ENF rapidly and efficiently, the CNN is designed to implement the extraction of deep spatial features, as shown in Figure 3.

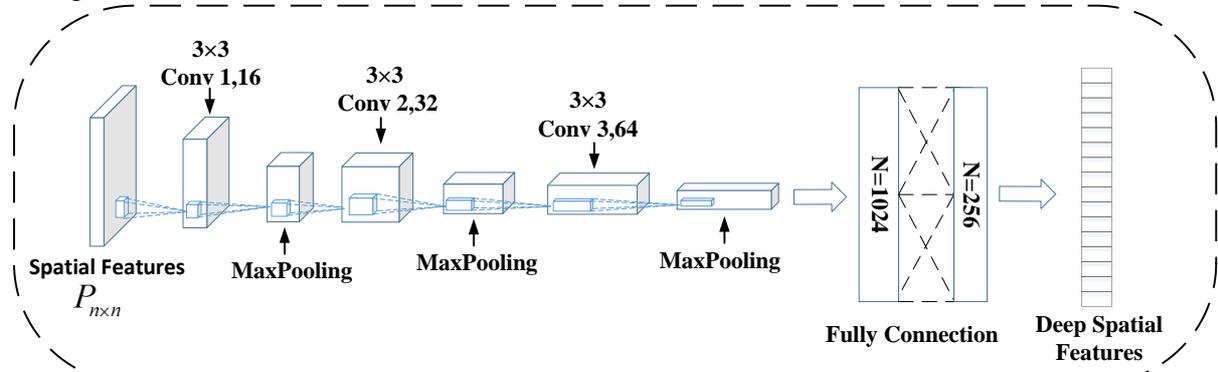

Figure 3: Deep spatial feature extraction module based on the CNN, the input is spatial features $P_{n \times n}$, $n$ is 45, and the output is the deep spatial features extracted by the CNN.

The CNN is a typical model of the deep neural network. The sparsity of convolutional kernel parameter sharing and inter-layer connections enables it to learn audio features with less computational effort and with stable results. In this paper, a deep spatial feature extraction module is constructed by taking advantage of the excellent spatial feature representation capability of CNN. The deep spatial feature extraction module consists of three convolutional layers, three pooling layers, and two fully connected layers, as shown in Figure 3. The input to the network is the spatial feature $P_{n \times n}$, which has a size of (45, 45). The number of convolutional kernels in the three convolutional layers is 16, 32, and 64, respectively, and the convolutional kernel size is (3, 3) for all three layers. The 3*3 convolutional kernel has more activation functions, richer features, and greater discriminative power than a larger convolutional kernel, which increases the non-linear expressiveness of the network and reduces the network parameters. Each convolutional layer is followed by a MaxPooling layer, both of size (2,2). The pooled high-level feature maps not only reduce the dimensionality and number of parameters of the original feature maps but also avoid problems such as over-fitting. The pooling equation is:

$$H = E(Y_\alpha) + b_1 \qquad (5)$$

where $Y_\alpha$ represents the original feature maps, matrix $E$ is the pooled domain of the feature maps, $b_1$ is the bias, and the pooled feature map $H$ is obtained by traversing the pooling domain of the original feature map. After three layers of convolutional pooling, the multi-dimensional data is transformed into one-dimensional data by a fully-connected layer. The numbers of nodes in the two fully-connected layers are 1024 and 256 respectively. The deep spatial feature extraction module extracts the local features of spatial features by the convolutional layer together with the pooling layer and then extracts the global features by the fully connected layers to finally obtain the deep spatial features.

### 3.2.2 Extraction of deep temporal features based on BiLSTM

Considering the fluctuation of ENF phase information over time, the deep temporal feature extraction module consists of two BiLSTM modules and two fully connected layers. BiLSTM consists of a forward LSTM and a backward LSTM, which can process the time series information in ENF from two different directions. Compared with LSTM, BiLSTM can better explore the correlation between the before and after time series data information in temporal features, and determine whether there are anomalies and mutations. The LSTM module controls information flow and avoids the long-term dependency problem of time series by adding one cell unit and three gate mechanisms, which consist of three gate structures: input gate, forget gate, and output gate. The combined action of these three gates allows the LSTM network to retain or discard previous state information to obtain more comprehensive sequence information. The relevant formulas for the LSTM are as follows:

$$\left.\begin{aligned} i_t &= \sigma(w_i \cdot [y_{t-1}, x_t] + b_i) \\ f_t &= \sigma(w_f \cdot [y_{t-1}, x_t] + b_f) \\ g_t &= tanh(w_g \cdot [y_{t-1}, x_t] + b_g) \\ c_t &= f_t * c_{t-1} + i_t * g_t \\ o_t &= \sigma(w_o \cdot [y_{t-1}, x_t] + b_o) \\ y_t &= o_t * tanh(c_t) \end{aligned}\right\} \quad (6)$$

where $i_t$, $f_t$, $o_t$ denote the input, forget and output gates, $w_f$, $w_i$, $w_o$ denote the weight matrix, $b_f$, $b_i$, $b_o$ denote the bias terms, and $y_t$ denotes the hidden state vector at time $t$.

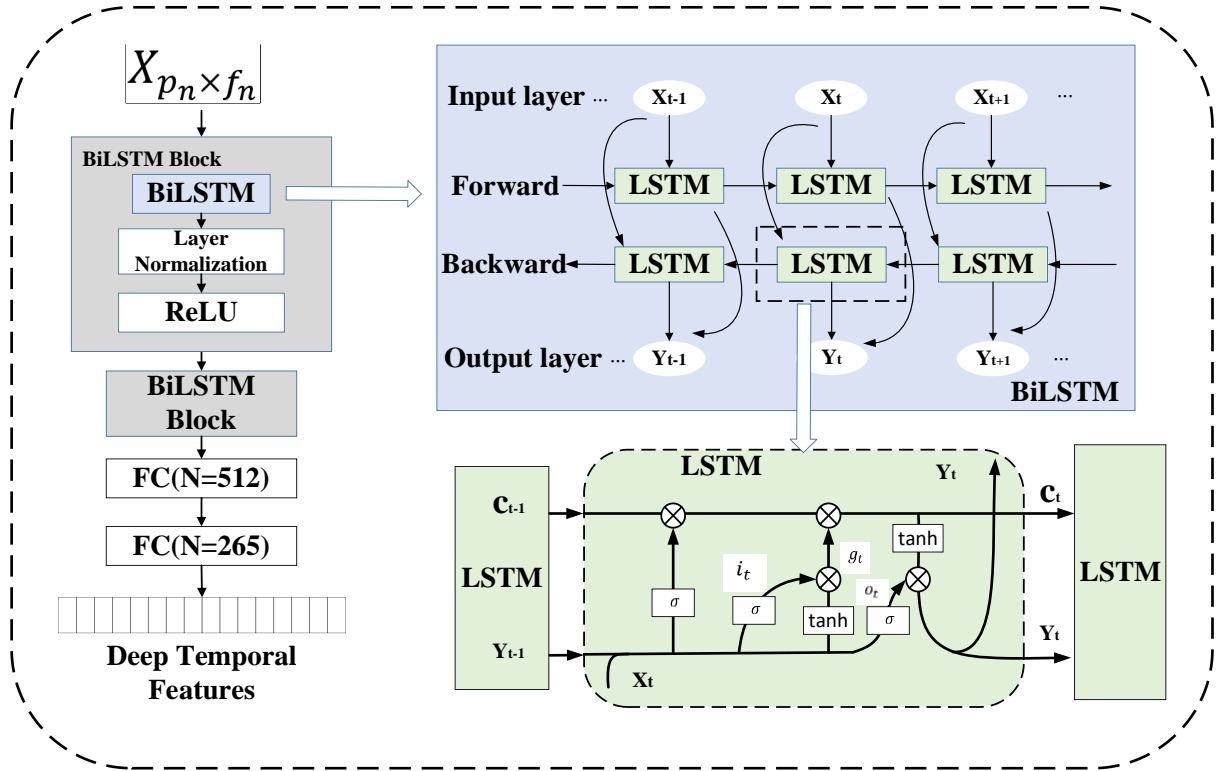

Figure 4: Deep temporal feature extraction module based on BiLSTM with the input of $X_{p_n \times f_n}$, $p_n$ of 85, $f_n$ of 25, and output of deep temporal features.

The deep temporal feature extraction module consists of two BiLSTM blocks and two fully connected layers, as shown in Figure 4. The input to the network is the temporal features $X_{p_n \times f_n}$, whose size is (85, 25). Each BiLSTM module includes a BiLSTM layer and a normalization layer, where the LSTM output dimension $unit$ is 85 and the normalization layer can improve the training speed and accuracy of the model and make the model more robust. The number of neurons in the two fully connected layers is 512 and 256 respectively. Finally, the output is deep temporal features.

### 3.2.3 Feature fusion networks based on attention mechanisms

To obtain spatio-temporal features with better representational power, this paper fuses deep spatial and temporal features based on an attention mechanism. Firstly, the deep spatial and temporal features are spliced to obtain the feature volume of length $L$. Then, a compression operation is applied to the feature volume of length $L$ by three fully connected layers to obtain more accurate weights. The activation functions of the fully connected layers are both ReLu and their number of neurons are $L$, $L/8$, and $L$ respectively. Next, the weights of the spatial and temporal features are obtained by connecting a fully connected layer with the number of neurons $L$. The activation function of the fully connected layer is Sigmoid [26]. Finally, the obtained weights are multiplied with the spliced feature volume to obtain the fused spatio-temporal features, as shown in Figure 5. The attention mechanism designed in this paper gives different weights to spatial and temporal features through automatic learning, and the features that have a high impact on the classification results are given greater weights to improve detection accuracy.

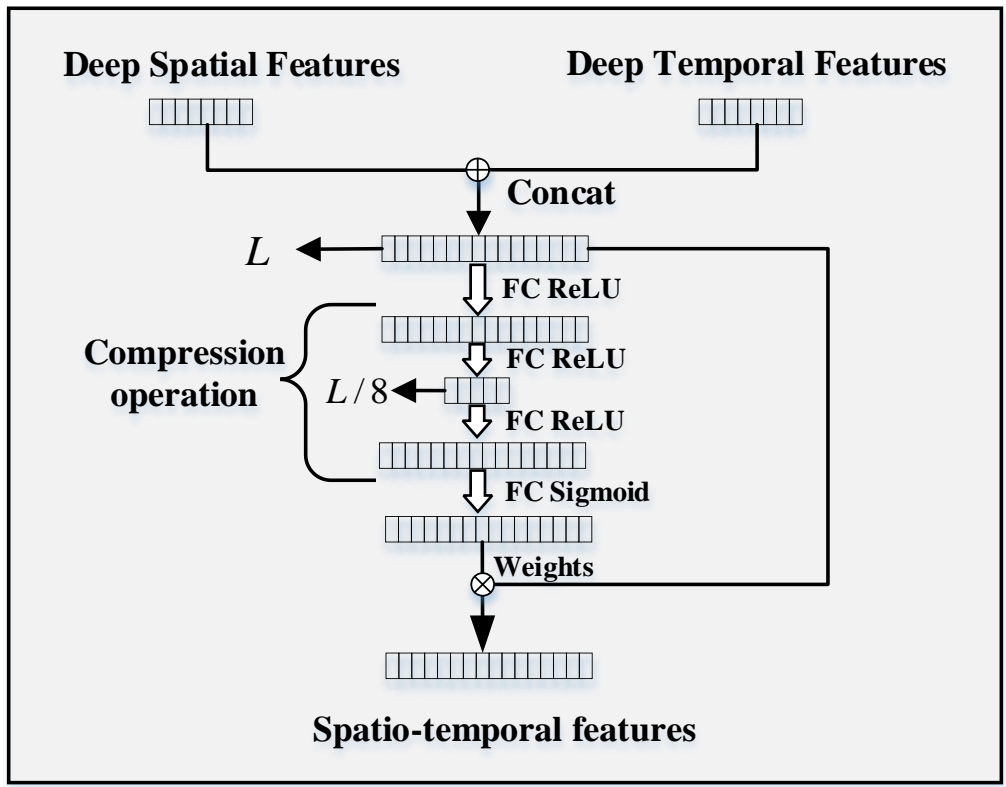

Figure 5: Deep spatio-temporal feature fusion based on attention mechanism.

### 3.2.4 Back-end classification network

The fusion feature with a high concentration of tampered information is obtained by the attention mechanism above. In this section, the MLP classification network is designed to determine whether it has been tampered with. The MLP network consists of four fully-connected layers (the number of neurons is 400, 256, 128, and 32, and the activation function is Leaky ReLU) and one softmax layer, with a dropout (dropout rate = 0.2) added between the layers to prevent overfitting, as shown in Figure 2. The softmax layer outputs a binary classification result for determining whether the audio has been tampered with, with the formula:

$$\hat{y}^{(i)} = soft\,max(x^{(i)}W + b) \tag{7}$$

where $\hat{y}^{(i)}$ is the predicted value of the output, $x^{(i)}$ is the output feature of the previous layer, $W$ is the weight and $b$ is the bias.

## 4. Experiments

All of our experiments are performed on two datasets: Carioca [17, 24] and New Spanish [25]. The Carioca database consists of Carioca 1 [17] and Carioca 2 [24], which contains voice signals from telephone recordings on the public switched telephone network (PSTN). The voice signals for the New Spanish database come from two public Spanish databases, AHUMADA and GAUDI. We conducted experiments on the Carioca, New Spanish, and Carioca & New Spanish datasets respectively, the details and divisions of the datasets are shown in Table 1.

Table 1: Data set information.

| The data set | Carioca | New Spanish | Carioca & New Spanish |
|---|---|---|---|
| Edited audio | 250 | 502 | 752 |
| Original audio | 250 | 251 | 501 |
| Total audio | 500 | 753 | 1253 |
| Audio time [s] | 9~35 | 16~35 | 9~35 |
| The training set | 319 | 480 | 800 |
| The validation set | 80 | 121 | 201 |
| The test set | 101 | 152 | 252 |

The experiments are implemented on the professional server with NVIDIA TITAN RTX high-performance graphics card and Tensorflow 2.0 development environment. The experimental parameters are as follows: In the spatial features $P_{n \times n}$, the number of phase points and the number of frames per unit frame are both 46; In the temporal features $X_{p_n \times f_n}$, the number of phase points per unit frame $p_n$ is 85 and the number of frames $f_n$ is 25; Loss function: binary cross-entropy; Optimiser: Adam; Training period: 300; Batch size: 64; Initial learning rate: 0.001.

In this section, the proposed model is validated on different datasets to demonstrate the robustness, reusability, and effectiveness of the model. We set up five experiments: (1) Comparison with existing work. (2) Analyze the effectiveness of different frame lengths during feature extraction. (3) Verify the effectiveness of the spatio-temporal features. (4) Verify the effectiveness of the attention mechanism feature in the fusion network. (5) Verify the effectiveness of the parallel CNN-BiLSTM network framework.

### 4.1 Comparison with existing work

To validate the effectiveness of the proposed digital audio tampering detection method based on spatio-temporal feature fusion, we compare the method proposed with existing work [17, 18] and our previous work [19] in the public Carioca and New Spanish datasets. DFT1-SVM [17] used $DFT^0$ and $DFT^1$ to estimate the phase and frequency of ENFC and SVM for classification. ESPRIT-SVM [18] used ESPRIT to estimate the features of phase peaks and SVM for classification. Our previous work [19] used $DFT^0$, $DFT^1$, and Hilbert transform to extract the phase and instantaneous frequencies of ENFC and used an optimized SVM for the classification of tamper detection. The results of the experiment are shown in Table 2.

Table 2: Comparison with existing work.

| Method | Accuracy [%] |
| --- | --- |
| DFT1-SVM [17] | 90.50 |
| ESPRIT- SVM [18] | 95.50 |
| previous work [19] | 92.60 |
| This work | 97.62 |

The results show that the proposed method improves the detection accuracy by 2.12%-7.12% over three state-of-the-art methods [17, 18, 19], outperforming the existing methods and verifying the effectiveness of the ENF-based spatio-temporal feature fusion method proposed in this paper.

**4.2 Analyze the effectiveness of different frame lengths during feature extraction**

The digital audio signal is a time-varying signal, where the various information and parameters characterizing the basic features of the audio change with time, but remain largely unchanged over a short period of time, so the audio signal has long-time volatility and short-time stability. Every audio clip of 0.17s contains ten phase points, and the phase number $p_n$ determines the amount of information about the ENF phase change in a short period of time. A suitable frame length setting can better represent the fluctuation of time-series information and is more conducive to the extraction of time-series information by the BiLSTM network. To analyze the effect of frame length on features, we conduct comparative experiments on the frame length settings of ENF temporal features on the Carioca, New Spanish, and Carioca &New Spanish datasets. Nine experimental groups are set up in the frame length range of 0.255s to 1.615s, and the experimental results are shown in Table 3. And the variation curve of accuracy with frame length is shown in Figure 6.

Table 3: Effect of different frame lengths on the experiment.

| Frame lengths | $p_n$ | $f_n$ | Carioca [%] | New Spanish [%] | Carioca & New Spanish [%] |
| --- | --- | --- | --- | --- | --- |
| 0.255s | 15 | 137 | 92.08 | 90.13 | 96.43 |
| 0.425s | 25 | 83 | 95.05 | 90.13 | 96.03 |
| 0.595s | 35 | 59 | 96.04 | 91.45 | 96.83 |
| 0.765s | 45 | 46 | 97.03 | 91.45 | 96.03 |
| 0.935s | 55 | 38 | 97.03 | 91.45 | 95.24 |
| 1.105s | 65 | 32 | 97.03 | 92.11 | 97.22 |
| 1.275s | 75 | 28 | 96.04 | 92.11 | 97.22 |
| 1.445s | 85 | 25 | 98.02 | 92.76 | 97.62 |
| 1.615s | 95 | 22 | 96.04 | 91.45 | 97.22 |

From the experimental results, it can be seen that high detection accuracy is achieved when the frame length is 0.595s and the number of phase points per unit frame $p_n$ is 35. When the frame length exceeds 1.445s and the number of phase points per unit frame $p_n$ exceeds 85, the accuracy tends to decrease. As the highest accuracy is achieved when the frame length is 1.445s and the number of phase points per unit frame $p_n$ is 85, we choose these two parameters and a number of frames of 25 as the parameters for the ENF temporal features $X_{p_n \times f_n}$.

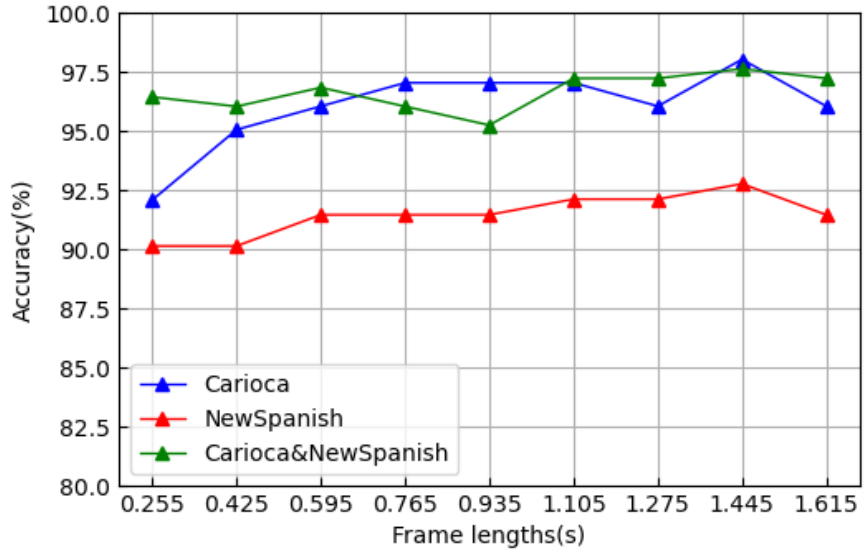

Figure 6: The variation curve of accuracy with frame length.

### 4.3 Verify the effectiveness of the spatio-temporal features

To verify the effectiveness of our proposed spatio-temporal feature, we select four features to compare and verify the ability of spatio-temporal features to represent tampered information, including the high-precision phase features $\varphi_{DFT^1}$ [17], the fusion feature $F_1F_2F_3$ [19], the spatial features $P_{46\times46}$ and the temporal features $X_{85\times25}$, where $P_{46\times46}$ and $X_{85\times25}$ were extracted in Section 3.1. Since SVM performs well in small sample classification tasks, we choose the SVM classifier for training and testing on $F_1F_2F_3$ and $\varphi_{DFT^1}$. Since CNN have excellent capability of spatial representation, we experiment with the spatial features $P_{46\times46}$ in a CNN model. Since BiLSTM networks handle temporal signals well, we perform experiments on the temporal features $X_{85\times25}$ in BiLSTM networks. We conduct experiments comparing the spatio-temporal features proposed in this paper with the above four features on the public dataset Carioca & New Spanish, and the experimental results are shown in Table 4.

Table 4: Compare with existing features.

| Feature | Model | Accuracy [%] |
| --- | --- | --- |
| $\varphi_{DFT^1}$ [17] | SVM | 90.48 |
| $F_1F_2F_3$ [19] | SVM | 96.03 |
| $X_{85\times25}$ | BiLSTM | 95.83 |
| $P_{46\times46}$ | CNN | 95.60 |
| Spatio-temporal features | CNN +BiLSTM | 97.60 |

As can be seen from Table 4, the accuracies of the $F_1F_2F_3$ and spatio-temporal features are higher than that of $\varphi_{DFT^1}$, $X_{85\times25}$, and $P_{46\times46}$, demonstrating the stronger ability of the fused features to represent tampered information. The accuracy of spatial features in CNN reaches 95.83% and that of temporal features in BiLSTM networks reaches 95.60%, indicating that the spatial and temporal features based on ENF have an excellent ability to characterize tampered information. The proposed spatio-temporal features full utilization of the static spatial information and time-varying information in the ENF phase and is higher than the temporal features by 2.00% and higher than the spatial features by 1.83% in detection accuracy. In summary, the proposed spatio-temporal features can obtain more tampering information in the ENF and has a stronger representation capability.

## 4.4 Verify the effectiveness of the attention mechanism in the fusion network

To verify the effect of the attention mechanism on parallel CNN-BiLSTM networks. We design a comparison experiment between the feature fusion network based on the splicing structure and the attention mechanism. The former uses a splicing layer to replace the attention mechanism module to achieve the fusion of deep spatial and temporal features, and the other network parameters remain unchanged.

Table 5: Verifying the effect of attentional mechanisms.

| Method | Carioca [%] | New Spanish [%] | Carioca & New Spanish [%] |
| --- | --- | --- | --- |
| This paper | 98.02 | 92.76 | 97.60 |
| No attention | 97.03 | 90.13 | 96.43 |

The experimental results are shown in Table 5, our parallel CNN-BiLSTM network based on the attention mechanism shows a significant improvement in accuracy over the splicing structure. This demonstrates that the attention mechanism is effective for different types of feature weight distributions. By means of adaptive learning, the attention mechanism can give more weight to features that are useful for classification and suppress invalid features, thus improving the detection accuracy.

## 4.5 Verify the effectiveness of the parallel CNN-BiLSTM network framework

To verify the effectiveness of the parallel CNN-BiLSTM network framework, this section combines networks that represent temporal and spatial features respectively, with six groups. The six spatio-temporal network combinations consist of one of DNN, CNN, and ResNet for spatial representation and one of LSTM, and BiLSTM for temporal representation. The MLP is used in the experimental recognition section. The experiments test the representation ability of various combinations on spatio-temporal features, and the results are shown in Table 6.

Table 6: Comparison with different deep networks

| Method | Carioca [%] | New Spanish [%] | Carioca & New Spanish [%] |
| --- | --- | --- | --- |
| DNN+LSTM | 94.06 | 89.47 | 96.03 |
| DNN+BiLSTM | 95.05 | 90.79 | 96.43 |
| Resnet+LSTM | 95.05 | 90.79 | 96.03 |
| Resnet+BiLSTM | 97.03 | 90.79 | 96.83 |
| CNN+LSTM | 95.05 | 91.45 | 96.83 |
| CNN+BiLSTM | 98.02 | 92.76 | 97.60 |

As can be seen in Table 6, the CNN shows higher accuracy than the DNN and ResNet networks for spatial representation and recognition. In terms of temporal representation and recognition, the BiLSTM network exhibits a higher accuracy rate than the LSTM network. In addition, the CNN+BiLSTM network obtains the best accuracy in all three datasets, Carioca, New Spanish, Carioca & New Spanish, reaching 98.02%, 92.76%, and 97.60% respectively, verifying the effectiveness of the parallel CNN-BiLSTM network based on spatio-temporal fusion.

**4.6 Discussion**

In this section, five sets of experiments are designed to verify the validity of the proposed method. In Experiment 4.1, we compare it with existing state-of-the-art work. The results show that the proposed method performs better in terms of feature characterization ability and network representation learning and recognition, and has improved accuracy. In Experiment 4.2, we find that different frame lengths affect temporal features, and we design feature fusion experiments with different frame lengths. The experimental results show that the highest accuracy is achieved for a frame length of 1.445s and a phase point $p_n$ of 85 per unit frame. In Experiment 4.3, we compare the proposed spatio-temporal features with $\varphi_{DFT^1}$, $F_1 F_2 F_3$, $P_{46\times46}$, and $X_{85\times25}$. The results show that the fused spatio-temporal features achieve the highest accuracy rate, verifying that the spatio-temporal features have stronger representation capability. In Experiment 4.4, we verify the effectiveness of the feature fusion network based on the attention mechanism by designing controlled experiments with the splicing structure and the attention mechanism. In Experiment 4.5, we conduct comparison experiments using six sets of parallel networks for temporal and spatial features. The results show that the accuracy of the proposed parallel CNN-BiLSTM network is higher than that of other parallel networks. In summary, the proposed digital audio tampering detection method based on ENF spatio-temporal feature representation learning outperforms existing state-of-the-art methods.

# 5. Conclusions

This paper presents a digital audio tampering detection method based on ENF spatio-temporal features representation learning. For the extraction of spatial and temporal features, firstly, this paper utilizes a high precision DFT analysis method to extract the phase sequences of the ENF. The unequal phase sequences are split into frames by adaptive frame shifting to obtain a matrix of the same size to represent the spatial features of the ENF. At the same time, the phase sequences are split into frames according to the ENF timing change information to represent the temporal features of the ENF. On the parallel CNN-BiLSTM network model, we use CNN, which excels in extracting spatial features, to further extract deep spatial features. The BiLSTM network, which is excellent at processing temporal signals, is used to further extract deep temporal features, and the fusion of spatio-temporal features is achieved by an attention mechanism. Finally, the MLP network is used to determine whether the digital audio has been tampered with. The experimental results show that the proposed method has high accuracy in both Carioca and New Spanish databases, outperforming existing methods. In future work, we will focus on extracting ENF features with greater representational power and more effective deep networks for application in digital audio tampering detection. In addition, we will use the extracted ENF features to achieve precise localization of digital audio tampering detection.

## Data Availability

The datasets used in this study can be obtained from the respective authors on reasonable request.

## Conflicts of Interest

The authors declare no conflicts of interest.

## Funding Statement

The research work of this paper were supported by the National Natural Science Foundation of China (No. 61901165, 62177022, 61501199), Collaborative Innovation Center for Informatization and Balanced Development of K-12 Education by MOE and Hubei Province (No. xtzd2021-005), and Self-determined Research Funds of CCNU from the Colleges' Basic Research and Operation of MOE (No. CCNU20ZT010).